# SYNTHESIS, STRUCTURAL AND PHOTO PHYSICAL PROPERTIES OF PEROVSKITE OXIDE $(KNbO_3)_{1-x}+(La_2NiMnO_6)_x$ FOR PHOTOVOLTAIC APPLICATION

Md Sariful Sheikh[1*], A. Dutta[2], T. K. Bhowmik[1], S. K. Ghosh[3], S. K. Rout[3], T. P. Sinha[1]
[1]Department of Physics, Bose Institute, 93/1, A. P. C. Road, Kolkata-700009, India
[2]Department of Condensed Matter Physics and Material Sciences, S. N. Bose National Centre for Basic Sciences, Block - JD, Sector - III, Salt Lake, Kolkata-700106, India
[3]Department of Physics, Birla Institute of Technology, Mesra, Ranchi-835215, India
*Corresponding Author: Phone No: +91 3323031194; Fax: (+91) 3323506790;
Email: sarifulsekh@gmail.com

ABSTRACT: Solid solutions of perovskite oxides $(KNbO_3)_{1-x}+(La_2NiMnO_6)_x$ (x=0, 0.1, 0.2 and 0.3) with a variation of band gap (1.33-3.6 eV) have been introduced as a promising photovoltaic absorber. The structural characterization of the prepared samples was carried out using X-ray diffraction (followed by Rietveld refinement) and Raman experiment. As the doping percentage of the monoclinic $La_2NiMnO_6$ in the solid-solution increases, the crystal structure of host $KNbO_3$ becomes more symmetric from orthorhombic to cubic. A large reduction in the particle size has also been observed in the solid solutions in comparison to the pure $KNbO_3$. The band gap (~ 1.33 eV) of the synthesized solid solution x=0.1 is found to be very close to the Shockley-Queisser band gap value of 1.34 eV, which suggests the promising photovoltaic possibility in this material. Photoluminescence (PL) emission spectra reveal a strong PL quenching in the solid-solutions in comparison to the $KNbO_3$. The overall structural and optical studies suggest the promising photovoltaic possibility in $KNbO_3$/ $La_2NiMnO_6$ solid solution.

## 1 INTRODUCTION

There is growing research interest to search for a narrow direct band gap solar cell absorber, that can be easily synthesized using the simple experimental techniques and low cost source materials [1-3]. In this regard, various multiferroic and ferroelectric perovskite oxides, like, $BiFeO_3$, $Pb(Zr.Ti)O_3$, $KNbO_3$ (KNO), etc., have shown their potential in the current photovoltaic research [4-7]. Ferroelectric domains in these materials produce a large internal electric field, which helps to separate the photo generated electron-hole pairs by dielectric screening and results in over band gap large photo voltage in multiferroic and ferroelectric solar cells. But, the band gap of these materials is too high to absorb the total visible region of the sun light, which restricts the photovoltaic performance of these materials [4-7]. Hence there is a significant research interest to search for a direct and narrow band gap multiferroic or ferroelectric solar cell absorber for optimal photovoltaic performance [7-11]. KNO, a widely studied ferroelectric material, have shown its potential in various light sensing semiconductor applications, like, photovoltaic, photo-catalytic dye-degradation, water splitting for $H_2$ production, etc [7, 9, 12, 13]. But like all other multiferroic and ferroelectric materials, band gap ($E_g$ ~ 3.6 eV) of KNO is too large to obtain the optimal photovoltaic performance [7]. Recently, significant researches have been focused on the band gap tuning of KNO for photovoltaic application [7, 9, 14, 15]. However, still no significant performance has been reported on its photovoltaic behavior, and thus a large research efforts is highly desired in order to improve its photovoltaic performance. In this regard, we have tuned the band gap and enhanced the photo physical properties of KNO for high photovoltaic performance.

We have doped the narrow band gap ferromagnetic semiconductor $La_2NiMnO_6$ (LNMO) within the KNO lattice in order to synthesize narrow band gap multiferroic semiconductor material for the photovoltaic application. LNMO is a widely studied ferromagnetic semiconductor with its Curie temperature ($T_c$ ~ 280 K) very close to the room temperature, narrow direct band gap ($E_g$ ~ 1.08 eV), magneto-dielectric and photovoltaic effect [16-18]. The combination of wide band gap ferroelectric KNO and narrow band gap LNMO, are expected to posses the desired narrow band gap with the intrinsic ferroelectric property of KNO. A detailed structural, magnetic and optical properties of the synthesized KNO/LNMO solid solutions have been performed in order to elucidate its photo physical properties for photovoltaic application.

## 2 EXPERIMENTAL DETAILS

### 2.1 Material synthesis

First, KNO/LNMO solid solutions was prepared using standard solid state reaction technique. Powders of $Nb_2O_5$, $K_2CO_3$, $La_2O_3$, $NiCO_3.2Ni(OH)_2.4H_2O$ and $MnO_2$ were taken in stoichiometric ratios and mixed homogenously by grinding in acetone medium for 8 hours. Finally, the mixtures were calcined at 1073 K for 5 hours in an alumina crucible. The heating process was repeated to complete the all phase formation. The doping percentage of LNMO in the solid solution was 0, 10, 20 and 30 % in molar ratio.

### 2.2 Characterization

The room temperature X-ray diffraction (XRD) pattern of the prepared samples was carried out using a powder X-ray diffractometer (Rigaku Miniflex II diffractometer). The Rietveld refinement of the obtained XRD data was performed by using the Fullprof code [19]. The Raman spectrum of the prepared samples was obtained at an excitation wavelength of 488 nm (He-Ar laser) using a Lab–RAM HR 800 (Jobin-Yvon) Raman spectrometer. The particle size of the sample KNO was obtained from the field effect scanning electron microscope (FESEM) image. The morphology of the samples x=0.1, 0.2 and 0.3 were determined using transmission electron microscope (TEM) images (FEI

Tecnai G2, 200 KV). The UV-visible absorption spectrum of the prepared samples was taken by a Shimadzu UV-VIS spectrometer. Photoluminescence (PL) emission spectra measurement were performed using a fluorescence/luminescence spectrophotometer (LS 55, Perkin Elmer). The calcined powders were pelletized into discs using polyvinyl alcohol as a binder and sintered at 1123 K for the study of magnetism. The vibrating sample magnetometer (VSM; Lakeshore) was used to study the magnetization in the prepared samples.

## 3 RESULTS AND DISCUSSION

### 3.1 Structure and morphology

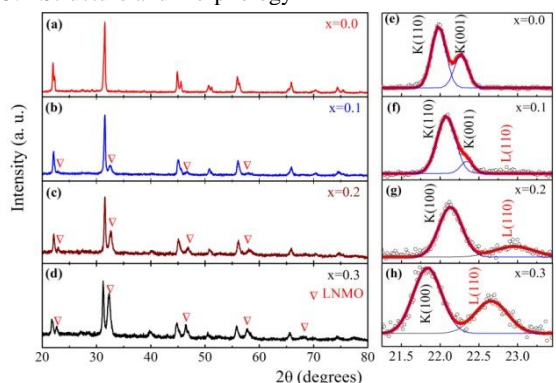

**Figure 1:** Room temperature XRD pattern of KNO/LNMO powder (a) x=0.0, (b) x=0.1, (c) x=0.2 and (d) x=0.3. Gaussian deconvolution of the XRD peaks of KNO/LNMO in the range from 21.3 to 23.5° of 2θ (e-h). The symbols, blue line, dark gray line and red line represent the experimental data, fit peak (KNO), fit peak (LNMO) and cumulative fit peak respectively. Suffix K and L represent the peaks originated from KNO and LNMO respectively.

The room temperature XRD patterns of the synthesized KNO/LNMO solid solutions are shown in Fig. 1 (a-d). From Fig. 1(a), it is observed that pure KNO sample crystallises in the orthorhombic phase with space group Bmm2 [JCPDS 04-007-9572]. The XRD peaks marked with the inverted delta symbol in Fig. 1(b-d) indicate the peaks originated from the monoclinic ($P2_1/n$) LNMO phase (JCPDS 04-013-0945 ). Fig. 1(e) displays the characteristic splitting of the (110) and (001) peaks for orthorhombic KNO. However it is clearly observed from Fig. 1(f-h) that in the sample x=0.1, this characteristic splitting of orthorhombic KNO reduces and disappears in the high doped KNO samples (x=0.2, 0.3). Disappearance of XRD peak splitting clearly indicates the recrystallization of KNO in the doped samples and points towards the possibility of more symmetric crystal structure of KNO. The XRD peaks of KNO in Fig. 1(c) matches very well with the tetragonal (P4mm) phase of KNO (JCPDS 71-0945). A significant shift of the (100) plane towards lower angle has been observed as shown in Fig. 1(d), which also indicates the possibility of recrystallization of KNO in the sample x=0.3. The XRD peaks of KNO in Fig. 1(d) may be due to the cubic (Pm-3m) phase of KNO [20]. Hence, to get a better understanding about the crystal structure of the solid solutions Rietveld refinement of the XRD data has been carried out and the corresponding refinement plots are shown in Fig. 2 (a-d). The values of the quality parameter ($\chi^2$) of refinement for the samples x=0.0, 0.1,0.2 and 0.3 are 1.6, 1.27, 1.38 and 1.72, respectively, which indicate the goodness of the refinement.

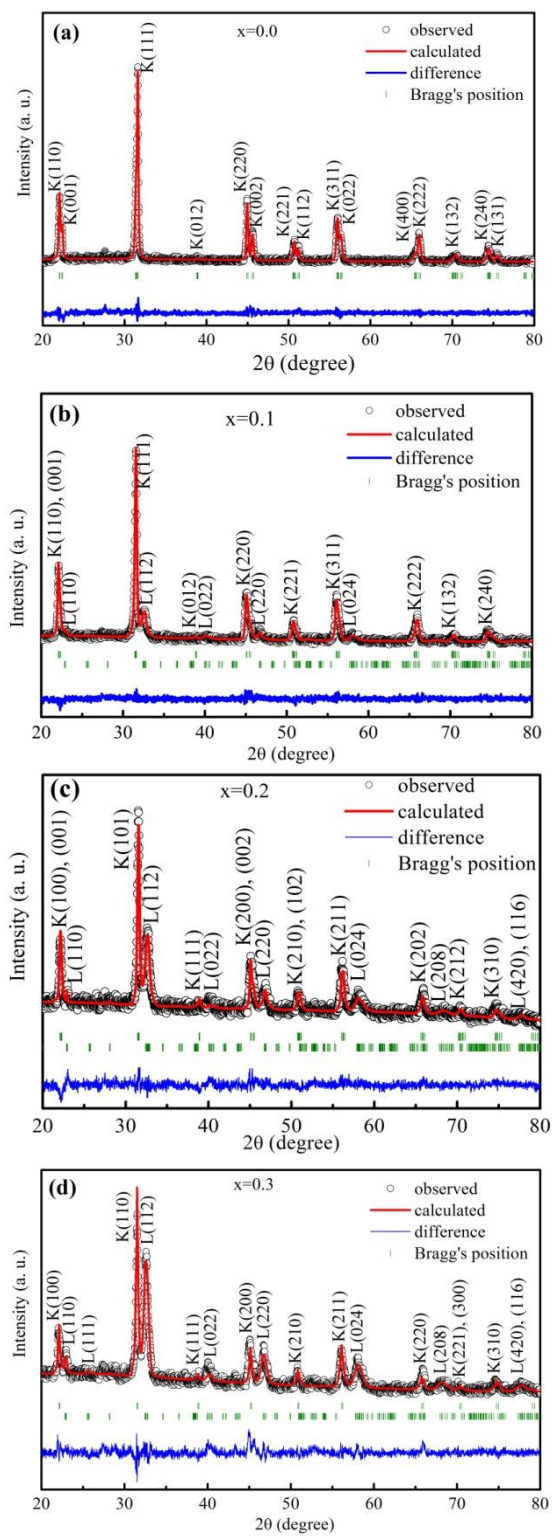

**Figure 2.** Rietveld refinement of the XRD data of solid solutions (a) x=0.0, (b) x=0.1, (c) x=0.2 and (d) x=0.3. Suffix K and L represent the crystal planes of KNO and LNMO lattice respectively.

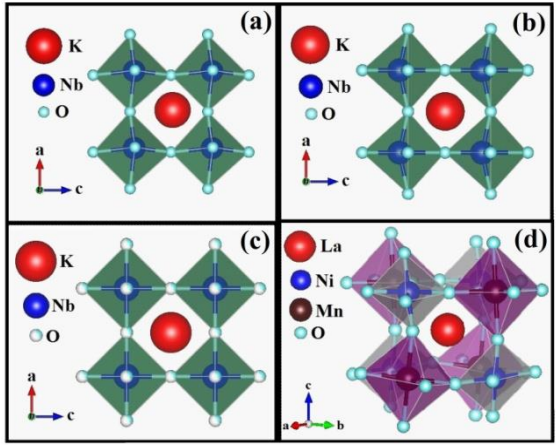

**Figure 3.** Unit cell representation of (a) orthorhombic, (b) tetragonal, (c) cubic KNO and (d) monoclinic LNMO.

The refined space groups and the lattice parameters of the samples are shown in Table 1, which are in good agreement with the previous reported lattice parameters of KNO and LNMO [20, 21]. It is clear that with the increase of LNMO percentage in the solid solution, the crystal structure of KNO becomes more symmetric from orthorhombic ($a \neq b \neq c$) to cubic ($a = b = c$). A schematic representation of the unit cells of KNO (orthorhombic, tetragonal and cubic) and LNMO (monoclinic) are shown in Fig. 3. Interestingly, due to the higher structural symmetry, cubic KNO shows better charge translation and photo reactivity than other less symmetric crystal structure of KNO [13]. Hence in respect of the crystal structure, KNO in the doped samples points towards the possibility of better charge transport as well as photovoltaic possibility in comparison to the pure KNO.

**Table I:** Refined lattice parameters

|        | KNO x=0.0 | KNO x=0.1 | KNO x=0.2 | KNO x=0.3 | LNMO x=0.3 |
|--------|-----------|-----------|-----------|-----------|------------|
| Phase  | Bmm2      | Bmm2      | P4mm      | Pm-3m     | P2$_1$/n   |
| a (Å)  | 5.705     | 5.684     | 4.019     | 4.013     | 5.503      |
| b (Å)  | 3.977     | 3.987     | 4.019     | 4.013     | 5.501      |
| c (Å)  | 5.698     | 5.689     | 3.992     | 4.013     | 7.747      |
| α      | 90°       | 90°       | 90°       | 90°       | 90°        |
| β      | 90°       | 90°       | 90°       | 90°       | 89.28°     |
| γ      | 90°       | 90°       | 90°       | 90°       | 90°        |

For a clear understanding about the structural properties of the solid solutions room temperature Raman spectra at an excitation of 488 *nm* has been carried out, which are shown in Fig. 4. In the orthorhombic (Bmm2) KNO phase (x=0.0) transverse and longitudinal phonon modes B$_1$(TO$_2$), B$_1$(TO$_4$), A$_1$(TO$_1$), B$_1$(TO$_3$), A$_1$(TO$_3$) and A$_1$(LO$_3$) are observed at 193, 278, 294, 532, 599 and 832 cm$^{-1}$, respectively [22].

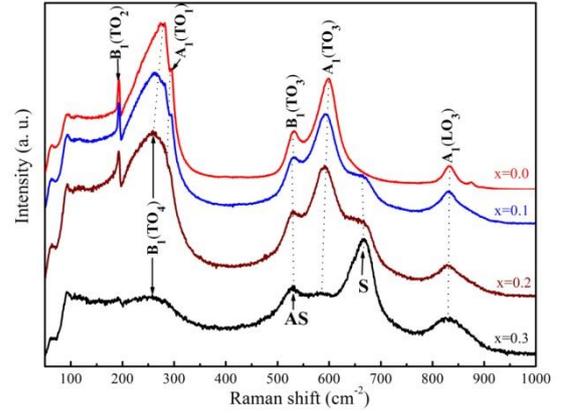

**Figure 4.** Raman Shift of the prepared samples.

But, in the doped samples the intensity of the lines B$_1$(TO$_2$) and A$_1$(TO$_1$) decreases as the doping percentage of LNMO increases. In the sample (x=0.3) the these two peaks become almost unobservable. This prominent change in B$_1$(TO$_2$) and A$_1$(TO$_1$) lines is related to the restructuring of the material and it represents the cubic phase of KNO in (x=0.3) [22]. In the cubic and tetragonal phase of KNO (x=0.2 and 0.3) the line B$_1$(TO$_3$) at 532 cm$^{-1}$ completely disappears. But in the doped samples (x= 0.2 and 0.3) a new line at 530 cm$^{-1}$ appears. This may arise due to the appearance of the anti-symmetric stretching (AS) mode vibration of the (Ni/Mn)O$_6$ octahedra at 530 cm$^{-1}$ [23]. The peak at 667 cm$^{-1}$ arises in the doped samples, which is related to symmetric stretching mode vibrations (S) of the (Ni/Mn)O$_6$ octahedra [23]. The intensity of the peak at 667 cm$^{-1}$ increases as the percentage of LNMO in the solid-solution increases. Hence, both the XRD and Raman experiment suggest the LNMO concentration dependent phase change in KNO host lattice.

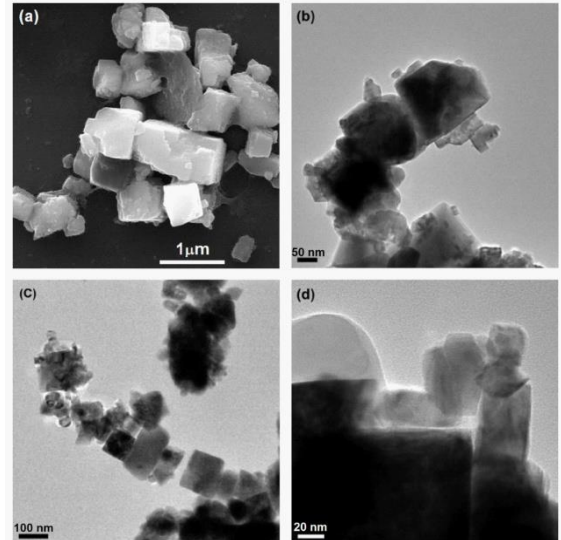

**Figure 5.** (a) FESEM image of KNO particle; TEM image of the sample (b) x=0.1, (c) x=0.2 and (d) x=0.3.

FESEM image as shown in Fig. 5(a) shows that the synthesized KNO particles are cubic in shape and the calculated average particle size is 465 *nm*. However, the TEM images of the samples x=0.1, 0.2 and 0.3 as shown in Fig. 5(b-d) denote a large reduction in the average particle size in the doped samples. The average particle size of the samples x=0.1, 0.2 and 0.3 are 41, 34 and 25 *nm*, respectively. Thus bulk to nano transformation has been occurred upon chemical doping. Interestingly, nanomaterials exhibit improved performance in various light sensing semiconductor applications due to their high surface area, increased surface energy and increased light absorption in comparison to the corresponding bulk materials [24, 25].

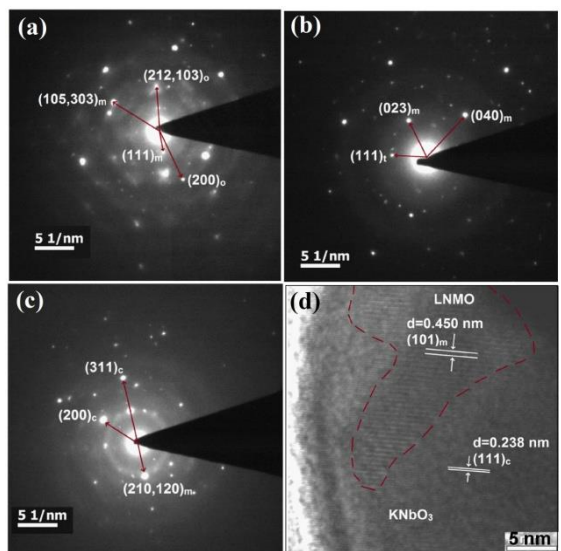

**Figure 6.** SAED pattern of sample (a) x=0.1, (b) x=0.2, (c) x=0.3; HRTEM image of sample (d) x=0.3. Prefix m, t, c and o represent the monoclinic, tetragonal, cubic and orthorhombic phases respectively.

Bright spots at regular positions in the selected area electron diffraction (SAED) patterns of the samples x=0.1, 0.2 and 0.3 as shown in Fig. 6(a-c) indicate the good crystalline nature of the doped samples. SAED patterns also confirm the coexistence of both KNO and LNMO lattice in the solid solutions. High resolution TEM (HRTEM) image of the sample x=0.3 as shown in Fig. 6(d) shows the formation of LNMO matrix within the KNO matrix. The large mismatch in the lattice parameters between KNO and LNMO, makes the LNMO matrix highly strained to accommodate within the KNO lattice and the crystal structure of KNO also changes from the orthorhombic to more symmetric cubic structure. The lattice misfit also prevents the host material KNO to form a larger lattice, which results a giant reduction in the average particle size of the doped samples in comparison to the pure KNO.

3.2 Magnetic property
Fig. 7 shows the temperature dependence of magnetic moment of the solid solutions in the temperature range from 80 *K* to 400 *K*. Though, KNO is a non-magnetic material, the solid solutions show ferromagnetic behavior and obviously, the magnetic contribution in the solid solutions originates from the ferromagnetic LNMO. Magnetic susceptibility ($\chi$) of a ferromagnetic material in the paramagnetic region is governed by the Curie-Weiss law: $\chi=C/(T-T_C)$, where $C$, $T$ and $T_C$ represent the Curie constant, measuring temperature and Curie temperature, respectively. Here, $T_C$ is estimated from the Curie-Weiss linear fit of the magnetic moment inverse ($M^{-1}$) versus temperature ($T$) plot (as shown in the inset of Fig. 7) in the paramagnetic region from temperature 250 to 350 K. The obtained value of $T_C$ using Curie-Weiss law is found to be 230, 236 and 242 *K* for sample x=0.1, x=0.2 and x=0.3, respectively. A gradual increment in $T_C$ has been observed as the LNMO concentration increases. Fig. 8 shows the magnetization versus field (*M-H*) curve at 80 *K*. The *M-H* curve as shown in Fig. 8, shows that the saturation magnetization also increases as the LNMO concentration in the doped samples increases and in all the cases magnetic moment tends to saturate above 5 *kG*, which indicates the ferromagnetic nature of the solid solutions. Inset of Fig. 8 shows that the remanant magnetization also increases with the LNMO doping percentage. However, coercivity almost remains the same in all the samples.

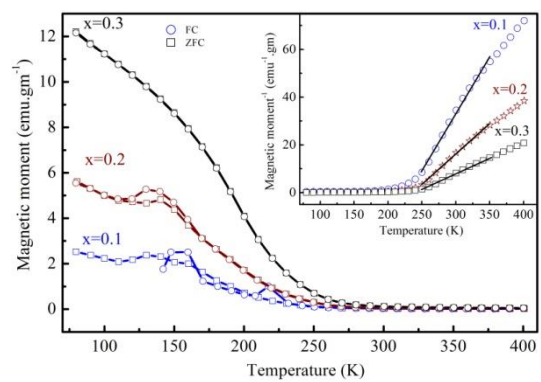

**Figure 7.** Temperature dependence of magnetic moment at a magnetic field of 2 *kG*. Inset shows the Curie-Weiss linear fit of $M^{-1}$ vs. T in the temperature range of 250 to 350 *K*.

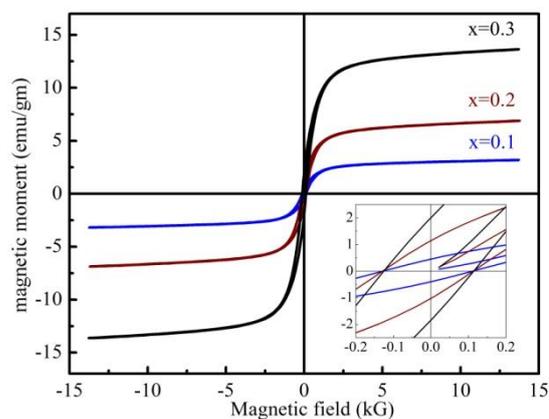

**Figure 8.** M-H loop of the KNO/LNMO samples at 80 K.

3.3 Optical property
The prime interest of this work was to reduce the band gap in the required range for the photovoltaic application of KNO as a promising light absorber. The optical band gap obtained from the Tauc plot as shown in Fig. 9 [26], is found to be 3.6, 1.33, 1.51 and 1.58 eV for the samples x=0.0, 0.1, 0.2 and 0.3, respectively. The band gap of the sample, x=0.1 matches very well with the Shockley-Queisser band gap value of 1.34 eV [27]. Shockley-Queisser band gap is the calculated value for which the theoretical power conversion efficiency of a single junction solar cell is normalized to the maximum value ( > 33 %) [27]. The band gap of another sample, x=0.2 (~1.51 eV) is also similar to that of $CH_3NH_3PbI_3$ (~ 1.5 eV), which has shown high photovoltaic power conversion efficiency in recent years(>20%) [28, 29]. However, over doping of LNMO in the KNO crystal lattice increases the band gap of the solid solutions. The synthesized solid solutions (x=0.1 and 0.2) are able to able to absorb the total ultraviolet and visible region of solar spectrum, which is essential for optimal photovoltaic performance. Hence in respect of the band gap, we can conclude that the newly designed solid-solution, $(KNO)_{1-x}+(LNMO)_x$ (x=0.1 and 0.2) can be used as a promising light absorbing material for the high performing solar cell fabrication.

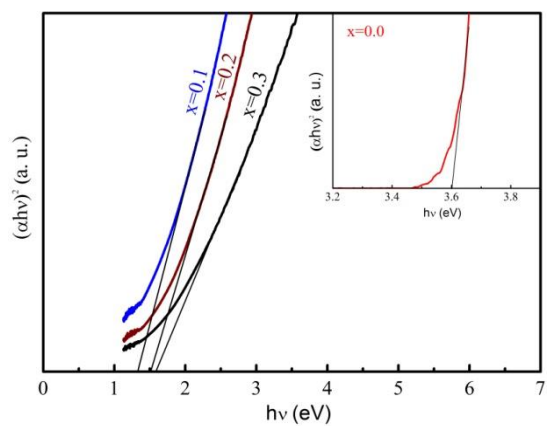

**Figure 9.** Tauc plot of the KNO/LNMO solid solution.

To further investigate the improved charge transport property in the materials, PL emission measurements were performed, which is shown in Fig. 10. It is observed that the doped samples present significant PL emission quenching in comparison to the pure KNO. Typically, a strong PL quenching of the light absorbing material provides an evidence of lower charge carrier recombination rate i.e. higher carrier life time in that material. Interestingly, high carrier life-time enhances the charge carrier extraction from the light absorber and hence improves the performance of a material in photovoltaic application. The enhanced charge carrier life time in KNO/LNMO may be due to the smaller particle size and better charge translation due to the more symmetrical crystal structure of KNO in the solid solutions.

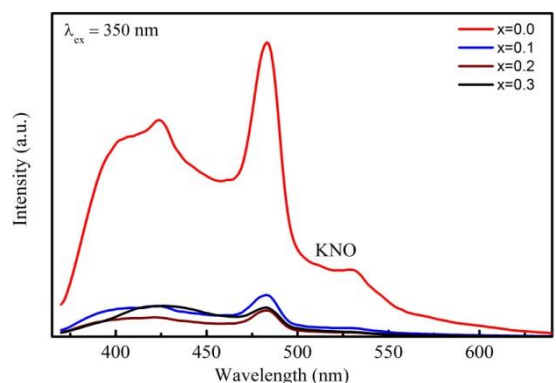

**Figure 10.** Photoluminescence emission spectra of KNO/LNMO solid solution.

## 4  CONCLUSION

A solid solution of KNO and LNMO has been introduced as a promising light absorber material for the photovoltaic application. The Rietveld refinement of the XRD data and Raman spectra of the synthesized materials show that with the increment in the LNMO doping percentage, the crystal structure of KNO becomes more symmetric from orthorhombic to cubic. Though KNO is a non-magnetic material, the doped samples show ferromagnetic behavior with high Curie temperature ($T_C \sim 242\ K$). The band gap of the solid solutions (x=0.1 and 0.3) is also very promising for high photovoltaic performance. A strong PL quenching confirms the lower carrier recombination rate and higher carrier life time in the doped samples in comparison to the KNO. Thus overall study shows that KNO/LNMO solid solutions posses many attributes suitable for high photovoltaic performance. KNO/LNMO solid solution may also find promising applications in photocatalytic dye degradation, water splitting, and other various light sensing semiconductor devices.

## 5  ACKNOWLEDGEMENTS

Md. S. Sheikh and T. K. Bhowmik would like to thank Department of Science and Technology (DST), Government of India for providing the financial support in the form of DST INSPIRE Fellowship (IF150220 and IF160418, respectively).